\documentclass[11pt]{article}
\usepackage{amsmath,amssymb,amsfonts,epsfig,hyperref,url}
\newcommand{\be}{\begin{equation}}
\newcommand{\ee}{\end{equation}}
\newcommand{\bea}{\setlength\arraycolsep{2pt} \begin{eqnarray}}
\newcommand{\eea}{\end{eqnarray}}
\newcommand{\nn}{\nonumber}
\def\ft#1#2{{\textstyle{\frac{\scriptstyle #1}{\scriptstyle #2} } }}

\begin{document}

\begin{flushright} {\footnotesize YITP-25-92, IPMU25-0031} \end{flushright}

\begin{center}
{\Large {\bf Constraining Cubic Curvature Corrections to General Relativity with Quasi-Periodic Oscillations}}

\vspace{20pt}

Alireza Allahyari$\, ^{1}$, Liang Ma$\, ^{2}$, Shinji Mukohyama$\, ^{3,4}$ and Yi Pang$\, ^{2}$

\vspace{10pt}

{\it $\, ^{1}$ Department of Astronomy and High Energy Physics, \\
Kharazmi University, 15719-14911, Tehran, Iran}

{\it $\, ^{2}$ Center for Joint Quantum Studies and Department of Physics,\\
School of Science, Tianjin University, Tianjin 300350, China }\\

{\it $\, ^{3}$ Center for Gravitational Physics and Quantum Information,\\
Yukawa Institute for Theoretical Physics,\\
Kyoto University, 606-8502, Kyoto, Japan}

{\it $\, ^{4}$ Kavli Institute for the Physics and Mathematics of the Universe (WPI),\\
The University of Tokyo Institutes for Advanced Study,\\
The University of Tokyo, Kashiwa, Chiba 277-8583, Japan}

\vspace{40pt}

\underline{ABSTRACT}
\end{center}

We investigate observational constraints on cubic curvature corrections to general relativity by analyzing quasi-periodic oscillations (QPOs) in accreting black hole systems. In particular, we study Kerr black hole solution corrected by cubic curvature terms parameterized by $\beta_5$ and $\beta_6$. While $\beta_6$ corresponds to a field-redefinition invariant structure, the $\beta_5$ term can in principle be removed via a field redefinition. Nonetheless, since we work in the frame where the accreting matter minimally couples to the metric, $\beta_5$ is in general present. Utilizing the corrected metric, we compute the QPO frequencies within the relativistic precession framework. Using observational data from GRO J1655$-$40 and a Bayesian analysis, we constrain the coupling parameters to $-12.31<\frac{\beta_5}{(5 M_\odot)^4}<24.15$ and $-1.99<\frac{\beta_6}{(5 M_\odot)^4}<0.30$ at 2-$\sigma$. These bounds improve upon existing constraints from big-bang nucleosynthesis and the speed of gravitational waves.

\thispagestyle{empty}
\pagebreak

\tableofcontents
\addtocontents{toc}{\protect\setcounter{tocdepth}{2}}

\section{Introduction}
In the framework of effective field theory, we can parameterize the action as the leading Einstein Hilbert term supplemented by  an infinite number of higher curvature terms suppressed by the scale $\Lambda_c$ above which new physics is needed to restore unitarity. In this spirit, a schematic form of the low energy effective action of quantum gravity is given by
\be
I=\frac1{16\pi G}\int d^4x\sqrt{-g}\left(R+\Lambda_c^{-2}{\cal L}_{4\partial}+\Lambda_c^{-4}{\cal L}_{6\partial}+\cdots\right).\label{genlag0}
\ee
In $D=4$, since the leading-order Kerr black holes are Ricci flat, i.e. $R_{\mu\nu}=0$, and the Gauss-Bonnet term is topological, the curvature squared terms will not modify the solution. Therefore the leading corrections to Kerr black hole come from cubic curvature interactions. In arbitrary spacetime dimensions ($D\geq8$), there are eight independent curvature cubed terms \cite{Fulling:1992vm}. However, in $D=4$, owing to the geometric identities (see Appendix~\ref{app:identities}),
\be
   E_6=R_{[\mu\nu}{}^{\mu\nu}
    R_{\rho\sigma}{}^{\rho\sigma}
    R_{\delta\gamma]}{}^{\delta\gamma}
    =0, \quad
    R_{[\mu\nu}{}^{\mu\nu}
    R_{\rho\sigma}{}^{\rho\sigma}
    R_{\gamma]}^{\gamma}
    =0,
\ee
the most general parity preserving cubic curvature action is parameterized by only six independent structures \footnote{Sometimes, in the literature (see \cite{Liu:2024atc} for an example), a different parameterization of curvature cubed structures is utilized in which $I_1=R^{\mu\nu}_{\ \ \lambda\rho}R^{\lambda\rho}_{\ \ \sigma\delta}R^{\sigma\delta}_{\ \ \mu\nu}$ corresponds to our $\beta_6$ term, and $I_2=R^{\mu\ \nu}_{\ \lambda\ \rho}R^{\lambda\ \rho}_{\ \sigma\ \delta}R^{\sigma\ \delta}_{\ \mu\ \nu}$ can be expressed as a combination of $\beta_1,\dots,\beta_5$ terms, using the identities given above. }
\bea
\mathcal{L}_{6\partial}&=&\beta_1R^3+\beta_2RR_{\mu\nu}R^{\mu\nu}+\beta_3R_\mu^\nu R_\nu^\lambda R_\lambda^\mu+\beta_4R_{\mu\lambda\nu\rho}R^{\mu\nu}R^{\lambda\rho}\cr
&&+\beta_5RR^{\mu\nu\rho\sigma}R_{\mu\nu\rho\sigma}+\beta_6R^{\mu\nu}_{\ \ \lambda\rho}R^{\lambda\rho}_{\ \ \sigma\delta}R^{\sigma\delta}_{\ \ \mu\nu}\ .
\label{R3}
\eea
Again the Ricci-flatness of the Kerr solution implies that only the last two terms in \eqref{R3} provide nontrivial corrections to the Kerr metric \cite{Ma:2023qqj,Bueno:2016ypa}. Within these two terms, only $\beta_6$ is invariant under field redefinitions, and thus it is the only term that contributes to black hole thermodynamics \cite{Ma:2023qqj} and multipole moments, as we shall see below \footnote{ In \cite{Glavan:2024cfs} an action-based order-reduction procedure was applied to the $\beta_6$ term to explicitly remove the spurious degrees from the EFT. The resulting effective action and equations of motion are expected to be useful e.g. for establishing a well-posed formulation of the numerical evolution of the system. }. We consider the most general parity even cubic curvature terms in the Lagrangian  because we assume that the matter still minimally couples to the gravity. If one performs a field redefinition to remove $\beta_1,\dots,\beta_5$ terms as considered in \cite{Cardoso:2018ptl, Cano:2019ore,  Maenaut:2024oci}, the matter coupling to gravity will have to contain higher curvature terms. In particular, in the  eikonal limit, a particle's trajectory will deviate from the geodesics. We are aware that theoretical work concerning infrared causality has imposed bounds on the cutoff scale suppressing the higher curvature corrections \cite{deRham:2021bll,Serra:2022pzl, Kehagias:2024yyp}, based on which whether EFT extensions of general relativity can have observational consequences has been questioned \footnote{The very recent work \cite{Alexander:2025gdn} suggested the opposite to the statement of  \cite{Serra:2022pzl}.}.
In this work, we shall adopt a more pragmatic viewpoint similar to \cite{Maenaut:2024oci}. Causality-based arguments reflect the difficulty to construct viable EFT of quantum gravity. However, it is still valuable to provide independent constraints on the EFT coefficients purely based on observations, given the fact that even km-scale corrections are not ruled out by current experiments \cite{Maenaut:2024oci}.

In this paper, we would also like to understand if the term proportional to $\beta_5$ affects other physical observables. The answer is not so straightforward given the fact that many physical quantities are measured through the metric which does receive corrections from the $\beta_5$ term.

We shall explore gravity in the strong-field regime by considering compact objects. A prospective approach to investigating  the nature of compact objects is provided by analyzing quasi-periodic oscillations (QPOs) detected in the X-ray emissions from accreting black holes in binary systems \cite{Abramowicz:2001bi, Pasham:2015tca, samimi}. These systems typically consist of a black hole or neutron star pulling matter from a nearby stellar companion. QPOs appear as distinct peaks in the power density spectra derived from X-ray light curves of such accreting objects \cite{Pasham:2018bkt, Belloni:2012sv}.

Several theoretical investigations have been put forward to interpret these narrow spectral features. Proposed mechanisms include relativistic precession, resonance phenomena, and $p$-mode oscillations within an accretion torus \cite{Stella:1997tc, Stella:1998mq, Stella:1999sj, Perez:1996ti,
Silbergleit:2000ck, Ingram:2020aki, Motta:2017sss, Bambi:2016iip,
Stuchlik:2008fy, Rezzolla:2003zx}. Each model seeks to explain the origin of the QPO signals and determine which physical processes are responsible. Among these, the relativistic precession model has gained considerable attention. It suggests that QPO frequencies correspond to the fundamental orbital frequencies of test particles moving around the central object \cite{miller, Maselli:2017kic,Franchini:2016yvq,Suvorov:2015yfv,Boshkayev:2015mna, Vahedi:2024hvr}. These particles exhibit small oscillatory motions when slightly displaced from their circular orbits. A key advantage of using QPOs as a probe is their high measurability, allowing for precise frequency determinations. Importantly, QPOs offer a unique opportunity to investigate the strong-field regime of gravity.

\section{Kerr black hole solution with higher-derivative corrections}
We apply the techniques developed in \cite{Cano:2019ore} to derive the perturbative corrections to the Kerr black hole. We expect the perturbative solution to preserve axisymmetry, so the metric ansatz takes the form
\bea
ds^2&=&-\Big(1-\frac{2\mu r}{\Sigma}-H_1\Big)dt^2-(1+H_2)\frac{4\mu ar(1-x^2)}{\Sigma}dtd\phi+(1+H_3)\Sigma\frac{dr^2}{\Delta}\cr
&&+(1+H_{5})\Sigma\frac{dx^2}{1-x^2}+(1+H_4)\Big(r^2+a^2+\frac{2\mu ra^2(1-x^2)}{\Sigma}\Big)(1-x^2)d\phi^2\ ,
\label{metricansatz}
\cr
\Sigma&=&r^2+a^2x^2,\qquad \Delta=r^2-2\mu r+a^2 \ .
\label{perturbed solution fix}
\eea
Here, \(H_i=H_i(r,x)\) depends only on $r$ and $x$, to be compatible with axisymmetry. Next, we expand $H_i$ in power series of the rotation parameter $\chi:=a/\mu$
\be
H_i(r,x)=\sum_{n=0}^\infty H_i^{(n)}(r,x)\chi^n\ .
\label{Hi}
\ee
Following \cite{Cano:2019ore}, $H_i^{(n)}(r,x)$ can
always be expressed as a polynomial in $x$ and in $1/r$
\be
H_i^{(n)}(r,x)=\sum_{p=0}^n\sum_{k=0}^{k_{\rm max}}H_i^{(n,p,k)}\frac{x^p}{r^k}\ ,
\label{Hin}
\ee
where $H_i^{(n,p,k)}$ are constants and for each undetermined function, the number of $k_{\rm max}$ depends on $n$ and $p$. Substituting \eqref{perturbed solution fix}, \eqref{Hi} and \eqref{Hin} into equations of motion, we solve for $H_i^{(n,p,k)}$ order by order in $\chi$ up to ${\cal O}(\chi^{10})$. In the computation below, we shall utilize the solution up to ${\cal O}(\chi^4)$, which is already quite lengthy. Thus we present the perturbative solution up to this order in an accompanying Mathematica notebook~\footnote{Available at \url{https://https://github.com/pangyi1/Kerr-Cubic}. In the solution, we have
chosen parameters $\mu$ and $\chi$ such that the mass and angular momentum do not depend on $\beta_5$ and $\beta_6$ at linear order. By contrast, the solution given in \cite{Cano:2019ore} did not make such a
choice. This is the main reason why our solution looks different from the solution there. }. In the data we will use, $\chi \sim 0.2-0.3$, which means our truncation of the perturbative solution at order ${\cal O}(\chi^4)$ should be good enough for the precision of the data.
We have chosen the parameters $\mu$ and $\chi$ properly so that up to linear order in $\beta_i$,  the mass and angular momentum remain uncorrected
\bea
M=\mu+\mathcal{O}(\beta_6^2),\quad J=\mu ^2 \chi+\mathcal{O}(\beta_6^2).
\eea
The radius of the black hole outer horizon is modified to be
\be
r_h=\mu+\mu  \sqrt{1-\chi ^2}-\frac{\beta _6}{\mu ^3}\Big(\frac{64}{231}+\frac{1555 \chi ^2}{3003}-\frac{616067 \chi ^4}{2586584}
\Big)+{\cal O}(\chi^6)\ .
\ee
Some formulas for thermodynamic quantities and unstable circular null orbits are shown in Appendices \ref{app:thermodynamics} and \ref{app:UCNO}, respectively.

\section{quasi-periodic oscillations}
In this section we apply relativistic precession model to find the constraints on the parameters $\beta_5$ and $\beta_6$.
Let us consider test particles orbiting around a stationary, axisymmetric source. The line element for this source is given by
\be
ds^2 = g_{tt} dt^2 + g_{rr}dr^2 + g_{\theta\theta} d\theta^2
+ 2g_{t\phi}dt d\phi + g_{\phi\phi}d\phi^2 \ ,
\ee
where $\theta$ is related to $x$ in (\ref{metricansatz}) as $x=\cos\theta$, and all the metric components do not depend on time and azimuthal angle $\phi$.
We have two constants of motion owing to the symmetries of the metric, the specific energy at infinity, $E$ and the $z$ component of the specific angular momentum $L_z$. We thus have
\bea
\dot{t} &= \frac{E g_{\phi\phi} + L_z g_{t\phi}}{
	g_{t\phi}^2 - g_{tt} g_{\phi\phi}} \,\label{eqt} , \quad \\
\dot{\phi}& = - \frac{E g_{t\phi} + L_z g_{tt}}{
	g_{t\phi}^2 - g_{tt} g_{\phi\phi}} \,\label{eqfi} \; ,
\eea
where the dot denotes derivative with respect to the affine parameter.
Using these constants in the normalization condition $g_{\mu\nu}\dot{x}^{\mu}\dot{x}^{\nu}=-1$, we obtain
\bea
g_{rr}\dot{r}^2 + g_{\theta\theta}\dot{\theta}^2 + V_{\rm eff}(r,\theta,E,L_z)
= -1\label{timelike} \, ,
\eea
where the effective potential is given by
\bea\label{eq:potential0}
V_{\rm eff} = -\frac{E^2 g_{\phi\phi} + 2 E L_z g_{t\phi} + L^2_z
	g_{tt}}{g_{t\phi}^2 - g_{tt} g_{\phi\phi}} \, .
\eea
According to the relativistic precession model, QPOs are related to the perturbations of circular orbits in the equatorial plane. So we consider circular orbits in the equatorial plane, $r=r_0, \theta=\pi/2$. By setting $\ddot{r}=\dot{r}=\dot{\theta}=0$ in the $r$-component of the geodesic equation we find that the angular frequency of circular orbits is given by
\bea \label{omega:fi}
\Omega_\phi =\frac{\dot{\phi}}{\dot{t}} =\frac{- \partial_r g_{t\phi}
	\pm \sqrt{\left(\partial_r g_{t\phi}\right)^2
		- \left(\partial_r g_{tt}\right) \left(\partial_r
		g_{\phi\phi}\right)}}{\partial_r g_{\phi\phi}} \, .
\eea
The energy and angular momentum for circular orbits are also given by
\bea
E &=& - \frac{g_{tt} + g_{t\phi}\Omega_\phi}{
	\sqrt{-g_{tt} - 2g_{t\phi}\Omega_\phi - g_{\phi\phi}\Omega^2_\phi}} \, , \\
L_z &=& \frac{g_{t\phi} + g_{\phi\phi}\Omega_\phi}{
	\sqrt{-g_{tt} - 2g_{t\phi}\Omega_\phi - g_{\phi\phi}\Omega^2_\phi}} \, .
\eea
Let us assume that the trajectory of the particle is subject to a slight perturbation in the  $r$ and $\theta$ directions, where we  have $r=r_0\left(1+\delta_r \right) $
and $\theta=\pi/2+\delta_{\theta}$.
The equation for the evolution of the perturbations is given by
\be
\label{eq:omegart}
\frac{d^2 \delta_r}{dt^2} + \Omega_r^2 \delta_r = 0, \quad
\frac{d^2 \delta_\theta}{dt^2} + \Omega_\theta^2 \delta_\theta = 0 \, ,
\ee
where we have
\be\label{eq:omegas}
   \Omega^2_r =  \left.\frac{1}{2 g_{rr} \dot{t}^2}
\frac{\partial^2 V_{\rm eff}}{\partial r^2}\right|_{r=r_0,\theta=\pi/2} \, , \quad
\Omega^2_\theta =  \left.\frac{1}{2 g_{\theta\theta} \dot{t}^2}
\frac{\partial^2 V_{\rm eff}}{\partial \theta^2}\right|_{r=r_0,\theta=\pi/2} \,.
\ee
According to relativistic precession model, QPOs are related to periastron precession
frequency  $\nu_p$ and nodal precession frequency $\nu_n$. They are defined as
\be
 \nu_p=\nu_\phi-\nu_r , \; \nu_n=\nu_\phi-\nu_\theta,
\ee
where $\nu_i=\Omega_i/2\pi$ with $i=\phi,r,\theta$.

We utilize the data available for GRO J1655$-$40 to constrain $\beta_5$ and $\beta_6$.
The detection of QPOs from this source was achieved through observations with RXTE \cite{Strohmayer, Remillard:1998ah}. Using X-ray timing techniques, three distinct QPO frequencies have been identified \cite{Motta:2013wga, Ingram:2020aki}. There are also other  X-ray binaries like XTE J1550$-$564. However, for this system two distinctive peaks are observed~\cite{Motta:2013wwa,Rink:2021mwt}. Our framework applies to X-ray binaries with three simultaneous peaks. The relevant data are summarized in table.\ref{table: data}.

To perform our Bayesian analysis, we define the $\chi$ square from QPOs as
\be
\label{eq:chi squred QPO}
\chi^2_{QPO}=\frac{(\nu_C -\nu_n)^2}{\sigma_C^2}+\frac{(\nu_L -\nu_p)^2}{\sigma_L^2}+\frac{(\nu_U -\nu_\phi)^2}{\sigma_U^2}\,,
\ee
where $\sigma_i$ ($i=C, L, U$) denotes the respective errors provided in Table~\ref{table: data}. As a result, the likelihood function takes the form $\mathcal{L}\sim e^{-\frac{1}{2}\chi^2_{QPO}}$.
\begin{table*}
    	\centering
\begin{tabular}{ccc}
   \hline
    & value&\\
   \hline
   	$\nu_U(Hz)$                 & $441\pm2$&   \\
   	$\nu_L(Hz)$                 & $298\pm4$ &     \\
   	$\nu_C(Hz)$         & $17.3\pm0.1$ &    \\
    \hline
\end{tabular}
	\caption{Parameters $\nu_U$, $\nu_L$ and $\nu_C$ for GRO J1655$-$40, where $\nu_C$, $\nu_L$ and $\nu_U$ are observed values for $\nu_n$, $\nu_p$ and $\nu_\phi$ respectively.
    The errors show 68\% confidence intervals obtained after fitting Lorentzian shapes to the power density spectrum using the \textit{XSPEC} package~\cite{Motta:2013wga}.
}
	\label{table: data}
\end{table*}


Additionally, we consider non-informative uniform priors as $0<a/M<1$, $4<r/M<10$, $-50<\beta_5/(5M_\odot)^4<50$ and $-50<\beta_6/(5M_\odot)^4<50$. The mass for this system has been determined from ellipsoidal light-curve fits to the quiescent B, V , R and I light curves~\cite{Beer:2001cg}. We use this uncertainty  assuming a Gaussian prior for the mass as
\be
\pi(M/M_\odot)\sim e^{-\frac{1}{2}\left(\frac{M/M_\odot-5.4}{0.3}\right)^2}\,.
\ee
The posterior is then given by

\be
\mathcal{P}  \propto \mathcal{L}\, \pi(M/M_\odot)\pi(a/M)\pi(r/M)\,\pi\Bigl(\beta_i/(5M_\odot)^4\Bigl),
\ee
where $i=5,6$.
\begin{figure}[htbp]
\centerline{\includegraphics[scale=0.60]{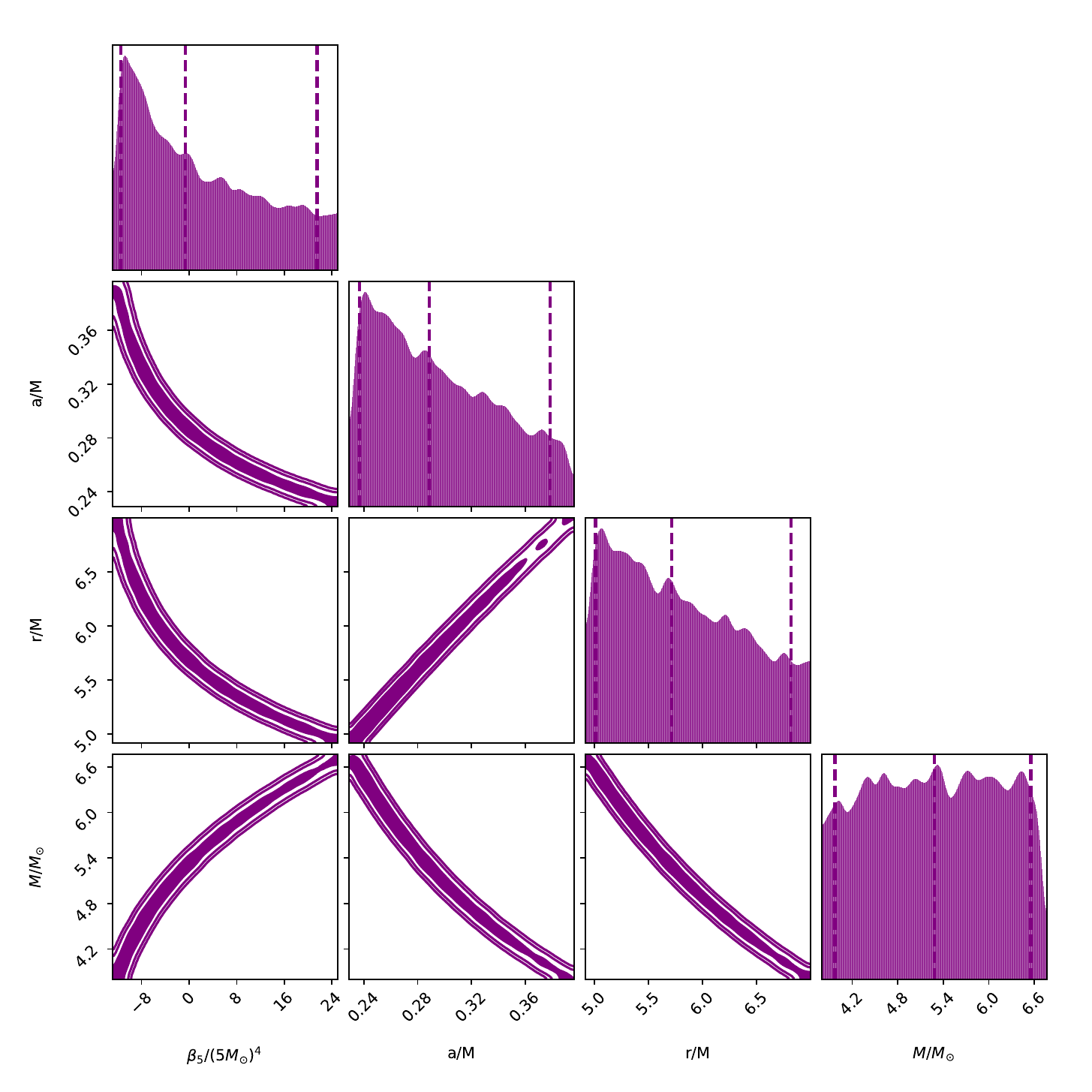}}
\caption{Two dimensional and one dimensional marginal distributions for $\beta_5$. The contours show 68\%, 95\% and 99\% credible intervals. Dashed lines are the 5\%, 50\%, and 95\% percentiles of the distribution. At 2-$\sigma$  we have $-12.31<\frac{\beta_5}{(5 M_\odot)^4}<24.15$. }
\label{fig1}
\end{figure}
\begin{figure}[htbp]
\centerline{\includegraphics[scale=0.60]{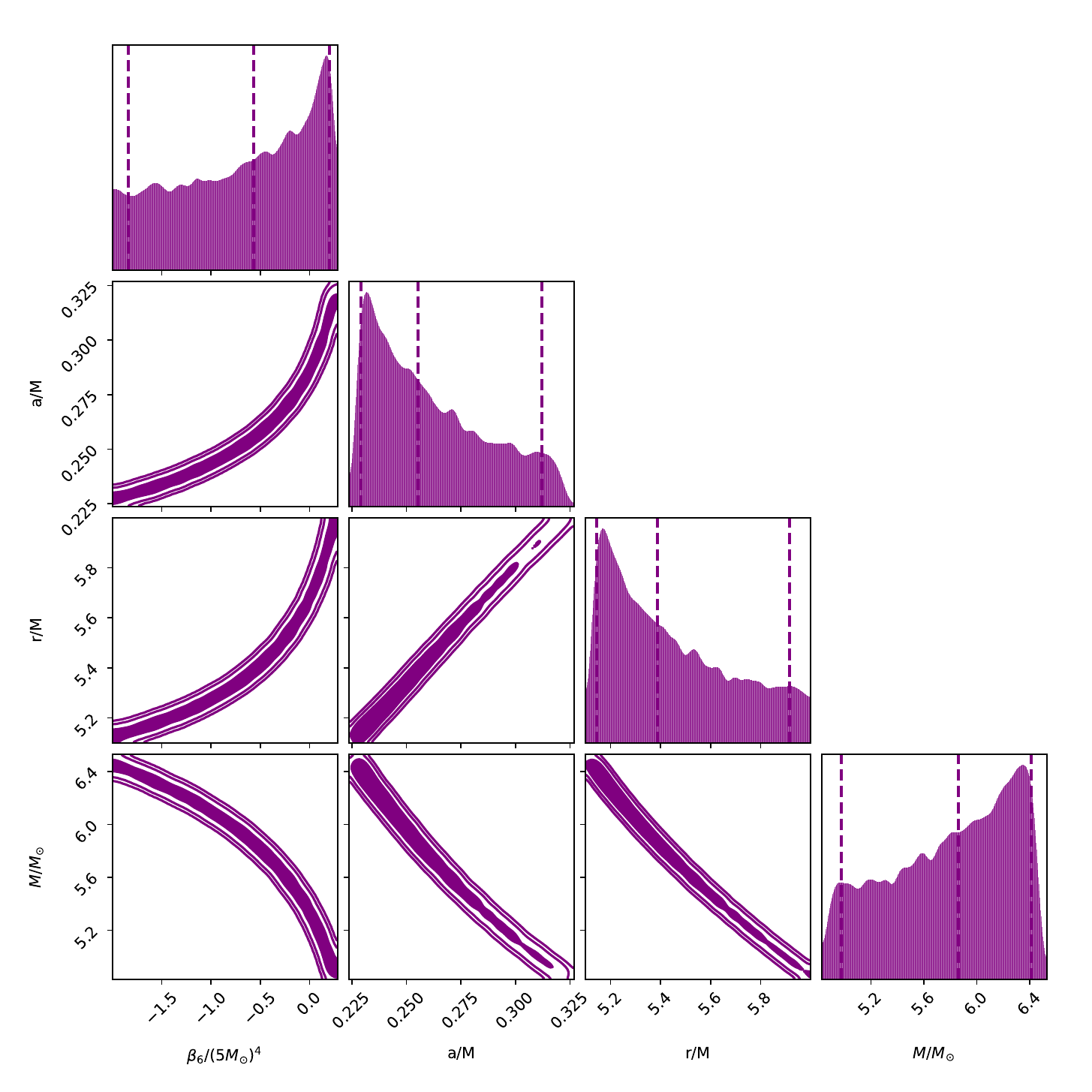}}
\caption{Two dimensional and one dimensional marginal distributions for $\beta_6$. The contours show 68\%, 95\% and 99\% credible intervals. Dashed lines are the 5\%, 50\%, and 95\% percentiles of the distribution. At 2-$\sigma$  we have $-1.99<\frac{\beta_6}{(5 M_\odot)^4}<0.30$.}
\label{fig2}
\end{figure}
The results are illustrated in figures~\ref{fig1} and~\ref{fig2} using \texttt{Dynesty}~\cite{Speagle:2019ivv}.
To run computations in parallel, we used the \texttt{pool} function in Dynesty to speed up the computations.
The plots show noticeable correlations between $\beta_5$ and other parameters such as $\chi$ and $M$. We also find that $\beta_6$, $\chi$ and $M$ are  highly correlated. This hinders our ability to constrain them with high precision. For $\beta_6$ we manage to find tighter bounds compared to $\beta_5$.
 We find that $-1.99<\frac{\beta_6}{(5 M_\odot)^4}<0.30$ at 2-$\sigma$ level.  For $\beta_5$ we have $-12.31<\frac{\beta_5}{(5 M_\odot)^4}<24.15$ at 2-$\sigma$ level.

\section{Discussion and Conclusion}
We have investigated a rotating black hole in a cubic curvature model where the Kerr solution in general relativity is  extended by two independent parameters. Our results show that both of the parameters enter the expressions for QPOs and lead to observational effects on QPOs. Our constraints put an upper bound on the suppression scale which is comparable to Schwarzschild radius of the compact object.

There are a number of other constraints on the suppression scale of the cubic curvature terms. From big-bang nucleosynthesis considerations we find that
\be
 \Lambda_c \gtrsim \frac{({\rm MeV})^2}{M_{\rm Pl}} \sim 10^{-24}{\rm GeV} \sim (10^5 {\rm km})^{-1}\,. \label{bound-nucleosynthesis}
\ee
 The curvature cubed corrections could also alter the speed of gravitational waves. By demanding that the correction to the propagation speed of gravitational waves should be less than about $10^{-15}$ for the present value of the Hubble expansion rate $H_0\sim 10^{-42}{\rm GeV}$, we obtain
\be
 \Lambda_c \gtrsim (10^{15}H_0^2\omega_{\rm LIGO}^2)^{1/4} \sim 10^{-28}{\rm GeV} \sim (10^9{\rm km})^{-1}\,. \label{bound-gravitationalwaves}
\ee
Our bound from QPOs is stronger than these bounds from the big-bang nucleosynthesis and the speed of gravitational waves. Our bound is also consistent with the causality bound \cite{deRham:2021bll} \footnote{The causality bound used here is derived from the requirement that the correction to the time delay caused by the higher curvature term is negative and unsolvable, namely $-\delta T>1/\omega$, where $\delta T$ is the correction to time delay, and $\omega$ is the frequence of the signal. In the recent work \cite{Alexander:2025gdn}, it is argued that the causality bound is not better than experimental constraints. There is no contradition between \cite{deRham:2021bll} and \cite{Alexander:2025gdn}, because the latter considered the modification to GR is exact rather than being perturbative correction and required the total time delay be negative. },
\be
\Lambda_c>7.42\times 10^{-19} {\rm GeV} (3M_\odot/M)\sim (0.1 {\rm km} )^{-1} (3M_\odot/M)\ .
\ee
It should be interesting to see, as we accumulate more experimental data for QPOs with improved precision in the future, whether the bound on the cutoff scale obtained from the data will move closer to the theoretical bound.

We have used the relativistic precession model to find the constraints on the parameters. However, there are also other effects that could contaminate the signal. These effects arise from the complexity of modeling the accreting material.  Moreover, the new parameters are correlated with other parameters like mass. This makes it difficult to constrain them more precisely.

Other  complementary studies such as study of the disk’s thermal spectrum and the broad $K\alpha$ iron line and gravitational waves could help find tighter bounds on the parameters~\cite{Reynolds:2002np,Fabian:2000nu}.

\section*{Acknowledgements}
This work of LM and YP is supported by the National Key Research and Development Program No. 2022YFE0134300, the National
Natural Science Foundation of China (NSFC) under Grant No. 12175164 and Tianjin University Self-Innovation Fund Extreme Basic Research Project Grant No. 2025XJ21-0007. L.M.~is also supported in part by National Natural Science Foundation of China (NSFC) grant No.~12447138, Postdoctoral Fellowship Program of CPSF Grant No.~GZC20241211 and the China Postdoctoral Science Foundation under Grant No.~2024M762338. The work of SM was supported in part by JSPS (Japan Society for the Promotion of Science) KAKENHI Grant No.\ JP24K07017 and World Premier International Research Center Initiative (WPI), MEXT, Japan.

\appendix
\section{Geometric identities and equations of motion} \label{app:identities}
In $D=4$, there are two identities built from cubic curvature terms
\bea
0&=&-12RR_{\mu\nu}R^{\mu\nu}+R^3+24R_{\mu\lambda\nu\rho}R^{\mu\nu}R^{\lambda\rho}+16R_\mu^\nu R_\nu^\lambda R_\lambda^\mu
-24R_{\mu\nu}R^{\mu\lambda\rho\sigma}R^\nu_{\ \lambda\rho\sigma}\cr
&&+3RR^{\mu\lambda\rho\sigma}R_{\mu\lambda\rho\sigma}+4R^{\mu\nu}_{\ \ \lambda\rho}R^{\lambda\rho}_{\ \ \sigma\delta}R^{\sigma\delta}_{\ \ \mu\nu}
-8R^{\mu\ \nu}_{\ \lambda\ \rho}R^{\lambda\ \rho}_{\ \sigma\ \delta}R^{\sigma\ \delta}_{\ \mu\ \nu}\ ,
\nn\\
0&=&4R_\mu^\nu R_\nu^\lambda R_\lambda^\mu-4RR_{\mu\nu}R^{\mu\nu}+\ft12 R^3+4R_{\mu\lambda\nu\rho}R^{\mu\nu}R^{\lambda\rho}+\ft12RR^{\mu\lambda\rho\sigma}R_{\mu\lambda\rho\sigma}
\nn\\
&&-2R_{\mu\nu}R^{\mu\lambda\rho\sigma}R^\nu_{\ \lambda\rho\sigma}\ ,
\eea
using which we reduce the independent cubic curvatures terms to those in \eqref{R3}.

The equation of motion used in our analysis is given as
\bea
R_{\mu\nu}-\frac{1}{2}g_{\mu\nu}R&=&
-P_{(\mu}^{\ \ \alpha\beta\gamma}R_{\nu)\alpha\beta\gamma}
-2\nabla^{\alpha}\nabla^{\beta}P_{(\mu|\alpha|\nu)\beta}
\nn\\
&+&\frac{1}{2}g_{\mu\nu}\left(\beta_5RR^{\mu\nu\rho\sigma}R_{\mu\nu\rho\sigma}+\beta_6R^{\mu\nu}_{\ \ \lambda\rho}R^{\lambda\rho}_{\ \ \sigma\delta}R^{\sigma\delta}_{\ \ \mu\nu}\right)%
\,,
\label{eom}
\eea
where we define the tensor $P_{\mu\nu\rho\sigma}$
\bea
P_{\mu\nu\rho\sigma}&=&\beta_5\Big[
\frac{1}{2}(g_{\mu\rho}g_{\nu\sigma}-g_{\mu\sigma}g_{\nu\rho})R_{\alpha\beta\gamma\lambda}R^{\alpha\beta\gamma\lambda}+2RR_{\mu\nu\rho\sigma}
\Big]+3\beta_6R_{\mu\nu}^{\ \ \alpha\beta}R_{\rho\sigma\alpha\beta}\,.
\eea
The right hand side of \eqref{eom} evaluated on the uncorrected Kerr black hole  serves as the effective stress tensor for the the perturbations at first order in $\beta_i$. The other terms in the effective Lagrangian \eqref{R3} are irrelevant in our discussion precisely because their corresponding effective stress tensor vanish on the uncorrected Kerr black hole solution.

\section{Thermodynamics of the corrected solution} \label{app:thermodynamics}
After fixing the mass and angular momentum, the corrections to other black hole thermodynamic quantities take the form up to ${\cal O}(\chi^{10})$
\bea
T&=&\frac{\sqrt{1-\chi ^2}}{4 \pi  \mu  (1+\sqrt{1-\chi ^2})}+\delta T+\mathcal{O}(\beta_6^2),\quad S=2 \pi  \mu ^2 (1+\sqrt{1-\chi ^2})+\delta S+\mathcal{O}(\beta_6^2),\cr
\Omega_H&=&\frac{\chi }{2 \mu  (\sqrt{1-\chi ^2}+1)}+\delta \Omega_H+\mathcal{O}(\beta_6^2)\ ,
\eea
where
\bea
\delta T&=&\frac{\beta _6}{64 \pi  \mu ^5}\Big(
1-\frac{7 \chi ^2}{2}-\frac{3 \chi ^4}{2}-\frac{7 \chi ^6}{8}-\frac{157 \chi ^8}{256}-\frac{1731 \chi ^{10}}{3584}
\Big)\ ,
\cr
\delta S&=&\frac{\pi  \beta _6}{2 \mu ^2}\Big(
1-\chi ^2-\frac{9 \chi ^4}{16}-\frac{11 \chi ^6}{28}-\frac{39 \chi ^8}{128}-\frac{225 \chi ^{10}}{896}
\Big)\ ,
\cr
\delta \Omega_H&=&\frac{5 \beta _6 \chi }{32 \mu ^5}\Big(
1+\frac{\chi ^2}{10}+\frac{3 \chi ^4}{70}+\frac{11 \chi ^6}{280}+\frac{377 \chi ^8}{8960}
\Big)\ .
\eea

Up to the same order, the mass and current multipole moments of the corrected black hole are given by
\bea
M_2&=&-\mu ^3 \chi ^2-\frac{6 \chi ^2 \beta _6 }{7 \mu }\Big(1-\frac{5 \chi ^2}{24}-\frac{\chi ^4}{16}-\frac{5 \chi ^6}{192}-\frac{5 \chi
   ^8}{384}\Big),\cr
M_4&=&\mu ^5 \chi ^4+\frac{242}{147} \beta _6 \mu  \chi ^4\Big(1-\frac{1241 \chi ^2}{5324}-\frac{1263 \chi ^4}{17303}-\frac{1587 \chi ^6}{50336}
\Big),\cr
M_6&=&-\mu ^7 \chi ^6-\frac{18448 \beta _6 \mu ^3 \chi ^6}{7623}\Big( 1-\frac{1647885 \chi ^2}{6715072}-\frac{1048245 \chi ^4}{13430144}
\Big),\cr
S_3&=&-\mu ^4 \chi ^3-\frac{69 \beta _6 \chi ^3}{98}\Big(
1-\frac{133 \chi ^2}{414}-\frac{637 \chi ^4}{6072}-\frac{665 \chi ^6}{14352}
\Big),\cr
S_5&=&\mu ^6 \chi ^5+\frac{629}{441} \beta _6 \mu ^2 \chi ^5\Big(1-\frac{2169 \chi ^2}{6919}-\frac{73599 \chi ^4}{719576}
\Big),\cr
S_7&=&-\mu ^8 \chi ^7-\frac{38855 \beta _6 \mu ^4 \chi ^7}{18018}\Big( 1-\frac{3432093 \chi ^2}{11112530}
\Big).
\eea

\section{Unstable circular null orbits} \label{app:UCNO}
In this appendix we find  unstable circular null orbits perturbatively, by using  null geodesic equations and $\dot{r}=\ddot{r}=0$. These orbits are important for  finding the shadow.
We have
\bea
   r_{0}&=& -\frac{16 a^4}{243 M^3}-\frac{5 a^3}{27 \sqrt{3} M^2}-\frac{2
   a^2}{9 M}-\frac{2 a}{\sqrt{3}}+\frac{\beta_5 X_5
   }{M^4}+\frac{\beta_6 X_6}{M^4}+3 M,\\
 X_5&=&-\frac{12880 a^4}{59049 M^3}-\frac{554 a^3}{2187 \sqrt{3}
   M^2}-\frac{176 a^2}{2187 M}-\frac{4 a}{81 \sqrt{3}}, \\
   X_6&=&\frac{448339059367 a^4}{6414878341872 M^3}-\frac{514616155
   a^3}{1655025372 \sqrt{3} M^2}-\frac{940061 a^2}{2786238
   M}-\frac{128300 a}{168399 \sqrt{3}}-\frac{2168 M}{6237}.
\eea
This expression reduces to  Schwarzschild result $r=3 M$ when $a=\beta_5=\beta_6=0$.
The important fact is when $a=0$, only $\beta_6$ appears in the expression for unstable null orbits.


\end{document}